\documentclass[12pt,preprint]{aastex}

\usepackage{amsmath}
\usepackage{amssymb}
\usepackage{latexsym}
\usepackage{subfig,graphicx}
\usepackage{txfonts}
\usepackage{natbib}

\bibpunct{(}{)}{,}{}{,}{,}

\begin{document}

\title{Observations of sausage modes in magnetic pores}

\author{R.J. Morton$^1$,  R. Erd\'{e}lyi$^1$, D.B. Jess$^2$ and M. Mathioudakis$^2$}

\affil{$^1$Solar Physics and Space Plasma Research Centre
(SP$^2$RC), University of Sheffield, Hicks Building, Hounsfield
Road, Sheffield S3 7RH, UK;\\
$^2$Astrophysics Research Center, School of Mathematics and Physics,
Queen's University, Belfast BT7 1NN, UK}

\email{[R.J.Morton,Robertus]@sheffield.ac.uk}

\begin{abstract} {We present here evidence for the observation of the
magneto-hydrodynamic (MHD) sausage modes in magnetic pores in the
solar photosphere. Further evidence for the omnipresent nature of
acoustic global modes is also found.} {The empirical decomposition
method of wave analysis is used to identify the oscillations
detected through a $4170$~{\AA} `blue continuum' filter observed
with the Rapid Oscillations in the Solar Atmosphere (ROSA)
instrument. Out of phase, periodic behavior in pore size and
intensity is used as an indicator of the presence of
magneto-acoustic sausage oscillations.} {Multiple signatures of the
magneto-acoustic sausage mode are found in a number of pores. The
periods range from as short as $30$~s up to $450$~s. A number of the
magneto-acoustic sausage mode oscillations found have periods of $3$
and $5$ minutes, similar to the acoustic global modes of the solar
interior. It is proposed that these global oscillations could be the
driver of the sausage type magneto-acoustic MHD wave modes in pores.
}{}
\end{abstract}

\keywords{plasmas - Sun: photosphere -  waves}

\shorttitle{Sausage modes in magnetic pores}
\shortauthors{Morton
et. al}

\maketitle

\section{Introduction}
The solar atmosphere is a highly dynamic magnetised plasma
containing a large array of distinct structures that are defined by
magnetic inhomogeneities. Each of these structures are able to
support a wide variety of oscillatory magneto-hydrodynamic (MHD)
modes. Over the last $50$ years there has been a steady increase in
the number of observations of oscillatory phenomena in the solar
atmosphere being reported (for latest reviews see, e.g.
\citealp{BANETAL2007}; \citealp{DEM2009}; \citealp{ZAQERD2009};
\citealp{MATetal2010}) showing waves and oscillations are ubiquitous
in the solar atmosphere. The increase in observations coincides with
the continuous improvement of space and ground-based observing
technology allowing unprecedented spatial and temporal resolution.

\begin{figure}
\includegraphics[scale=0.58,clip=true,viewport=0.5cm 1.0cm 14cm 14cm]{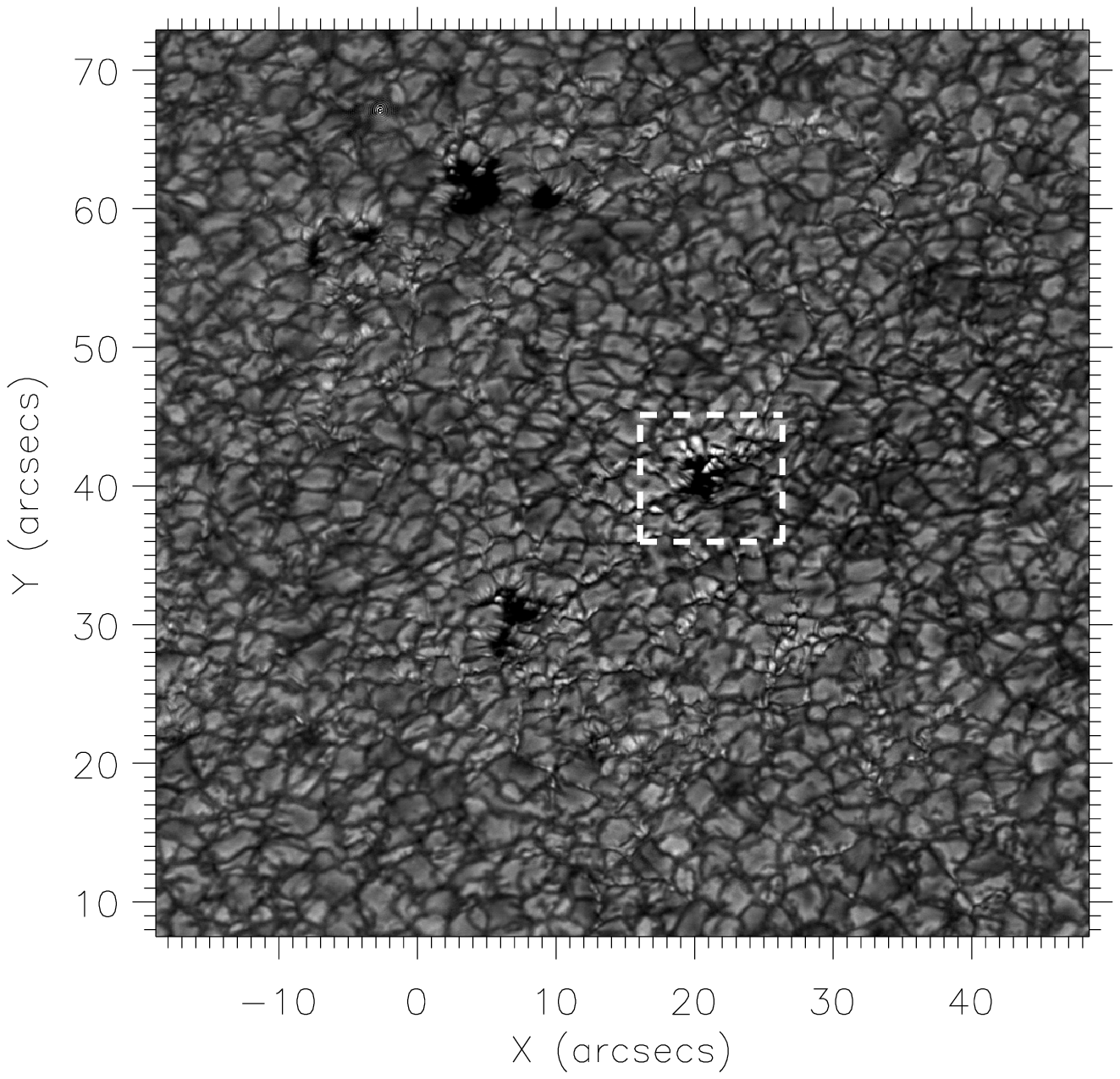}
\caption{Field of view observed by ROSA on $22$ August $2008$. Boxed
areas highlight pore in which oscillations are present.
}\label{fig:fov}
\end{figure}

This high spatial and temporal resolution permits detailed
investigation of some of the Sun's smallest, currently detectable
magnetic features. One such feature are magnetic pores that have
diameters ranging from $1-6$~Mm. The pores are regions of intense
magnetic fields ($\sim1700$~G) first visible in the photosphere and
expanding as they reach chromospheric heights. Pores show highly
dynamic behavior due to a constant buffeting from the convective
motion of granules at the photospheric level (\citealp{SOB2003};
\citealp{SANRIM2003}). There has recently been observations of
photospheric structures experiencing a vortex style motion which
could act as a driver for a wide variety of waves and oscillations
(\citealp{BONetal2008}). These waves and oscillations may be able to
propagate upwards through the various layers in the lower solar
atmosphere along the length of the pore, which itself acts as an MHD
waveguide. The majority of these waves will be reflected at the
transition region due to the steep gradients in sound or Alfv\'{e}n
speeds, the minority, however, will make it into the corona. The
transmitted portion of the waves may be relevant for MHD wave
heating or magneto-seismology of the solar corona (see, e.g.
\citealp{KLI2006}; \citealp{TARERD2009}). One of the exciting new
discoveries associated with magnetic elements in the lower solar
atmosphere is evidence for torsional Alfv\'{e}n waves
(\citealp{JESSetal2009}) which has, historically, been difficult to
detect. Would pores be able to support the other magneto-acoustic
modes? If yes, would these modes be sausage or kink? We answer these
questions in the current Letter.

The nature of MHD waves in sunspots and pores has been extensively
investigated, where the sunspots are modelled as thin,
gravitationally stratified flux tubes (\citealp{ROBWEB1978};
\citealp{ROB1981b}; \citealp{ROB1992}). The seminal theory of
\cite{EDWROB1983} describing waves in a straight, magnetic cylinder,
which has been applied extensively to coronal oscillations, was also
derived for photospheric conditions. This last aspect is somewhat
neglected and has only received limited attention although offering
equally rich physics and opportunity for solar magneto-seismology as
its coronal application. Dispersion diagrams for the photosphere
(see, e.g. \citealp{EDWROB1983}; \citealp{ERDFED2010}) show clearly
the nature of the waves supported by a photospheric waveguide.
Perhaps more relevant to waves in pores are the dispersion diagrams
presented in \cite{EVAROB1990}.

 The sausage oscillation is thought to be driven by the granular buffeting and
the vortex motion, so pores are a good candidate for the observation
of this oscillatory mode because of their compact structure. A
further postulated driver maybe \emph{p}-modes or magneto-acoustic
waves which propagate within the solar interior
(\citealp{DORetal2008}). The main feature of the sausage mode is the
periodic fluctuation of the cross-sectional area of the wave guide
(see, e.g. \citealp{EDWROB1983} for cylindrical and
\citealp{ERDMOR2009} for elliptical wave guides). Neither Alfv\'{e}n
nor kink oscillations would show evidence of perturbations of the
cross-section of the waveguide. The change in the area of the
cross-section caused by the sausage motion is also associated with
periodic fluctuations in density and temperature within the
waveguide.

Up until recently it has been almost impossible to find evidence for
periodic change in cross-sections of waveguides (in any layer of the
solar atmosphere), which indicates the presence of sausage
oscillations. This is due to the limitations of the spatial
resolution of many observing instruments. A number of earlier
observations have, however, attempted to identify the signature of
sausage oscillations via the indirect detection of intensity
oscillations, e.g using Doppler shifts (\citealp{TARetal2007};
\citealp{ERDTAR2008}) or periodicities in x-ray emission (e.g.
\citealp{NAKetal2003}) in the solar corona. To our knowledge, the
first reported periodic oscillations in pore size were by
\cite{DORetal2008}, who observed periods from $20$ minutes to $70$
minutes in a photospheric pore with the Swedish Solar Telescope
(SST). However, no intensity information was provided.

We present here observations of sausage oscillations in a pore
detected using the Rapid Oscillations in the Solar Atmosphere (ROSA)
instrument. The observations are in both pore size and intensity in
the best cases and solely in pore size in other cases. The periods
of the oscillations range between $50$~s and $600$~s and do not have
constant frequency suggesting continual evolution of the pore due to
dynamic behavior.

\section{Observations and Data Reduction}
One new and exciting ground-based setup is the ROSA instrument which
provides high spatial ($100$~km, $2$-pixel) and temporal ($0.03$~s)
resolutions. ROSA is situated at the Dunn Solar Telescope and
further details of the experimental setup and operation of ROSA are
given in \cite{JESetal2010}. ROSA is able to observe at multiple
wavelengths allowing the investigation of the magnetic connection
between the various layers of the lower solar atmosphere from deep
photosphere to the upper chromosphere.

\subsection{Data}
The data used here was taken by ROSA at ${15:24}$~UT on the $22$
August $2008$. A group of $5$ magnetic pores, shown in
Fig.~\ref{fig:fov}, were observed continually by ROSA for around
$1$~hr and $7$~mins with a $50$~{\AA} wide filter centered at
$4170$~{\AA}. The pores were formed before the observations began
and all are still present in the final image of the observing run.
The $4170$~{\AA} blue continuum filter samples the lower
photosphere. The spatial sampling is $0.069$'' per pixel, giving a
$2$-pixel spatial resolution of 0.138'' (or $100$~km) with an
overall field of view of $50200\times50100$~km$^2$. The cadence
obtained for this filter during the ROSA observing sequence was
$0.2$~s.
  The initial observations were processed through the ROSA data
reduction pipeline which removed dark current, read-out noise,
camera inconsistencies and variable light levels across the incident
beam. To improve image quality the speckle reconstruction method was
used with a ratio of ${64:1}$. After processing, the cadence of the
observations was reduced to $12.8$~s. Intensities are normalised to
the mean background value.

\begin{figure*}[htp]
\includegraphics[height=3.05cm, width=18cm]{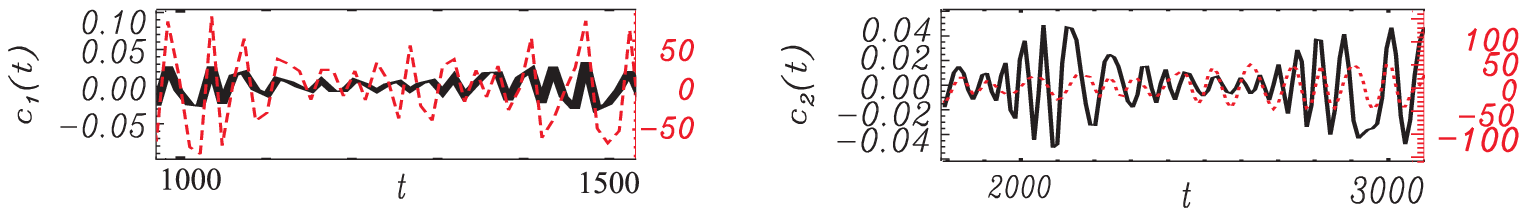}
\hspace{1cm}
\includegraphics[height=3.35cm, width=18cm]{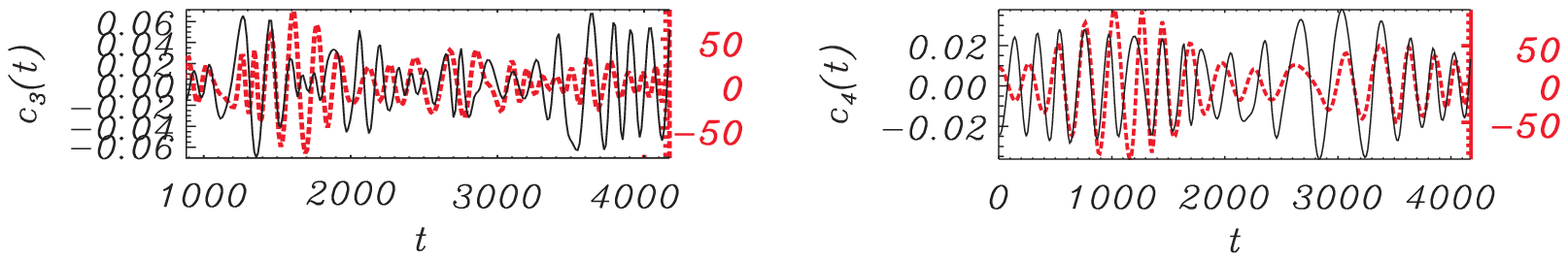}
\hspace{1cm}
\includegraphics[height=3.35cm, width=8.5cm]{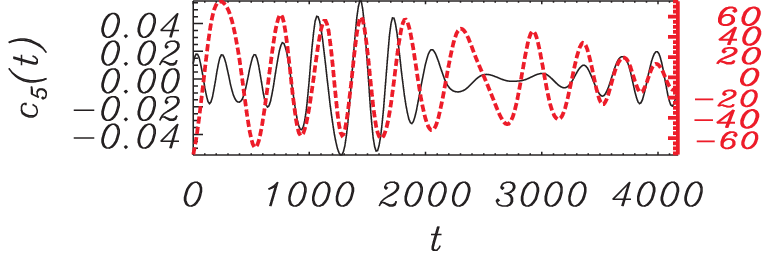}
\hspace{1cm}
\includegraphics[height=3.35cm, width=8.5cm]{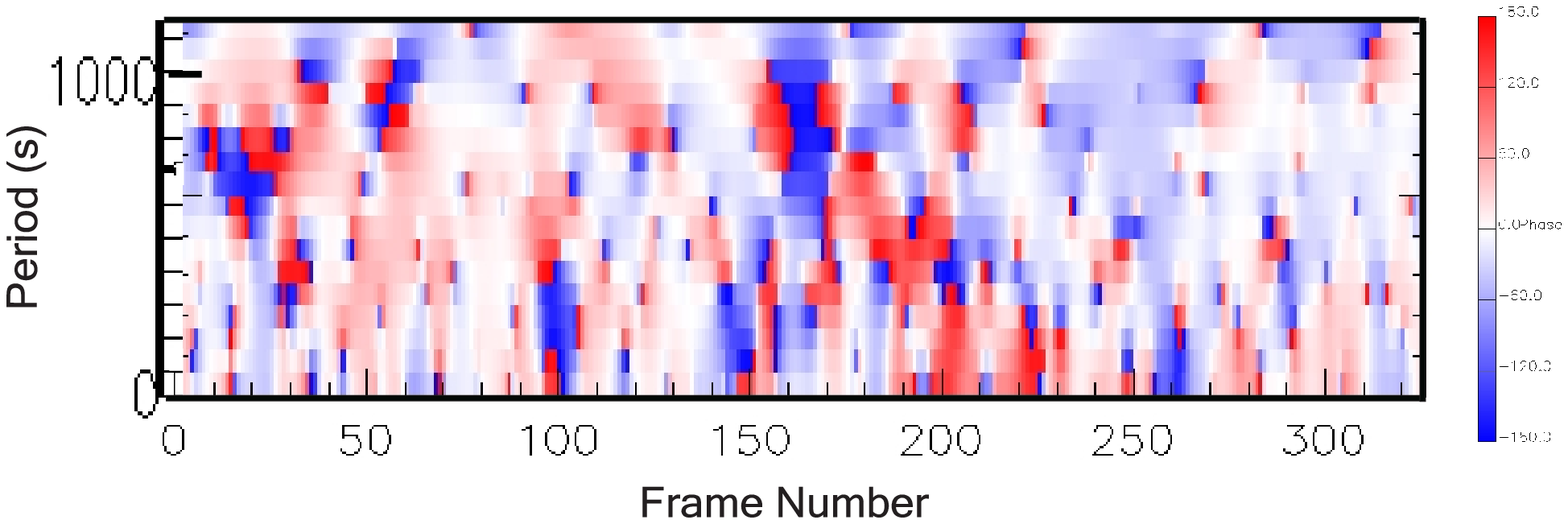}
\caption{Shown is a direct comparison between IMF components of
intensity (black solid lines) and pore size (red dashed lines). The
intensity is normalised with respect to the background mean value
and the pore size is measured in pixels. Last panel shows wavelet
phase plot for two the time series.}\label{fig:obs1}
\end{figure*}

\subsection{Analysis}\label{sec:analysis}
The pores are observed close to disk center, and assuming that a
pore is a waveguide which extends upwards into the solar atmosphere
with a base in the photosphere, the line of sight is near
perpendicular to the cross-section of the waveguide. This is ideal
when searching for the periodic fluctuation in the area of the cross
section. Images of the pore show it is distinctly non-circular in
cross-section and movies of the magnetic pore reveal that it is
highly dynamic. At present, there is no such theory describing
waveguides with a complicated, non-symmetric geometry and dynamic
behavior. Waves in more complicated models should still retain the
same basic properties (see, e.g. \citealp{RUD2003}) so the
approximation of a circular cross-section provides an adequate
representation of the pore. The periodic change in the area of the
cross-section is also accompanied by a periodic variation in
intensity, which should be $180$ degrees out of phase with the
change in cross-sectional area for the sausage mode.

The method used for obtaining the pore size is as follows. Firstly
we find the median value and variance of the intensity of the whole
field of view in each frame. A 'box' is placed around each of the
pores, large enough that the pore is contained within the box at all
times. The area of the pore is then defined as being the number of
pixels with a value of intensity $3\sigma$ less than the median
value of intensity, providing a $99\%$ confidence level that the
dark pore pixels are contoured.

The data representing the size of the pore is then analysed using
Empirical Mode Decomposition (EMD). This is a powerful tool that is
capable of resolving non-stationary and non-linear time series. The
theory is described in \cite{HUAetal1998} with excellent examples of
applications to solar phenomena given in \cite{TERetal2004}. EMD can
overcome some of the problems associated with other analysis methods
(e.g. wavelet analysis) and is suitable to decompose the time series
into a finite number of Intrinsic Mode Functions (IMFs). The IMFs
represent the different timescales of variations in the original
times series. To determine whether a sausage oscillation is present,
a direct comparison between IMFs of intensity and pore size with
similar timescales is performed. This shows clearly out of phase
behavior. A distinct sign of sausage oscillations is when periodic
phenomena in cross-section and intensity are almost $180$ degrees
out of phase. We shall refer to such a signal as a \emph{strong}
signal. Another, although not quite as distinct, signal is when
periodicities in pore size do not match with any intensity
variations. In such a case we can only assume the expected 'out of
phase' intensity signal has been hidden by another effect that
modifies the intensity. This will be referred to a \emph{weak}
signal in the following analysis. All signals will also have to be
longer than $\sqrt{2}P$ to be considered as a real oscillation. The
phase information is important for finding out of phase intensity
and pore size oscillations. Unless comparisons can be made between
the individual time series of intensity and pore size to compare the
phases, then it is virtually impossible to identify whether the
periodic oscillations in pore size are due to background intensity
changes or the sausage mode. The EMD method allows an inspection of
the different time-scales associated with the initial time-series
and proves an ideal tool for analysis. We, however, apply the
wavelet method to produce phase diagrams and compare to EMD results
as control.

\section{Oscillations in pores}
The pore under consideration is located in the dashed box centered
at ($15,48$)~arcsecs in Fig.~\ref{fig:fov}. The area is calculated
with a confidence value of $3\sigma$, i.e $99\%$ confidence. The
average size of the pore is $536$ pixels, which is roughly
$1.36\times10^6$~km$^2$.

The IMFs are found for the average intensity per unit area and pore
size time series and both time series reduce to $8$ IMF components
plus the trend. The IMFs have approximate characteristic periods
$<100$~s, $100$~s, $150$~s, $250$~s, $400$~s for IMFs $c_1$ to
$c_5$. They have also been compared to wavelet plots of the two time
series and the IMF components have periods that agree well with the
periods that have significant power in the wavelets.

\begin{table}[htp] \caption{Periods of identified oscillations}
\centering
 \begin{tabular}{ l c c c r  }
    \hline
    \hline
    IMF & Strength & Period & Start time & Duration \\
    & & (s) & (s) & (s) \\
    \hline
    $c_1$ & Strong & $30\pm 13$ & $1200$ & $100$  \\
    $c_2$ & Weak & $102\pm 30$ & $1900$ & $800$  \\
    $c_3$ & Weak & $140\pm 13$ & $1700$ & $400$  \\
     & Strong & $134\pm 13$ & $2100$  & $250$  \\
     & Weak & $126\pm 40$ & $2350$ & $1800$  \\
    $c_4$ & Weak & $180\pm 13$ & $0$ & $375$   \\
     & Weak & $281\pm 18$ & $1700$ & $750$   \\
    $c_5$ & Weak & $447\pm 13$ & $1600$ & $1350$   \\
          \hline
     \end{tabular}
  \label{tab:oscill}
\end{table}

In Fig.~\ref{fig:obs1} (top panels) we show a comparison between the
IMF components $c_1$ and $c_2$ of pore size (red dashed lines) and
average intensity per unit area (solid black lines) for a small
section of the time series. Fig.~\ref{fig:obs1} also shows a
comparison of the $c_3$, $c_4$ and $c_5$ components of intensity and
pore size for the entire time series. The different oscillatory
signals found are summarised in Table. \ref{tab:oscill}. The
oscillations here have an amplitude of $20-60$ pixels which
corresponds to a $5\times10^4-1.5\times10^5$~km$^2$ increase in
area, which is around $4-12\%$ of the average pore area. The small
amplitude of the oscillation suggests that the sausage waves are
linear in nature. To confirm the evidence for the sausage
oscillations found in the comparison of the IMF components, we
compare the two times series using cross wavelet analysis (see, e.g.
\citealp{GRIetal2004}). We calculate the phase difference of the two
time series in Fig.~\ref{fig:obs1} (bottom right panel). It can be
seen that the areas showing the strongest out of phase behavior
correspond to the periods and events identified in the IMFs.

\section{Which Mode}
When it comes to determining which mode each period in Table~$1$
corresponds to, i.e. fast or slow, we require more information. This
is due to the plasma-$\beta$ being close to $1$ and the periods of
the fast/slow modes are not as distinct as, say, in the corona. Due
to the lack of velocity information we are also unable to determine
whether the observed oscillatory signatures are standing or
propagating modes. Assuming the pore is a straight, finite flux tube
with a \textit{uniform cross-section} situated between the
photosphere and the transition region, then standing modes can be
set up with the transition region acting as a reflector (see, e.g.
\citealp{MALERD2007M}). We can use the periods obtained from the EMD
analysis to estimate the length of the waveguide that supports these
oscillations. Typical values for pores are $T=4000$~K, $B=2$~kG,
$\rho=10^{-8}-10^{-7}$~gcm$^{-3}$. These give values of
$v_A=B/\sqrt(\mu_0\rho)=17-56$~kms$^{-1}$ and $c_s=\sqrt(\gamma
T/\tilde{{\mu}})=9$~kms$^{-1}$ where $\mu_0=4\pi$, $\gamma=5/3$ and
$\tilde{\mu}=0.6$. The phase speed of the slow mode is close to the
tube speed which has estimated value, $c_T=8-9$~kms$^{-1}$. The
fundamental period is given by $P\approx2L/c_{ph}$ where $L$ is the
tube length. Assuming the fundamental period is given by the
oscillation identified in the $c_5$ IMF, which is $447\pm13$~s the
length of the tube is then calculated to be $1.9-2.1$~Mm. If we now
take the oscillation identified in the $c_4$ IMF, which has period
of $180$~s, and assume it is due to be the fast mode, we obtain
$L=1.5-5.0$~Mm for $c_{ph}=v_A=17-56$~kms$^{-1}$. Both these
estimates can provide a tube length which is comparable to the
widely accepted and commonly used height of the transition region
above the solar surface, i.e. required tube length to set up
standing mode oscillations, which is around $\sim2$~Mm. Note, this
is not full proof of the nature of the observed oscillations. If
future observations can identify the standing modes, then this
should help in fast/slow mode identification.

A comment should be made here: the assumption that the tube is
straight is a highly ideal assumption. The cross-section of the flux
tube is actually expected to be expanding with height. Also,
gravitational stratification plays a significant role in determining
the density structuring in flux tubes located the photosphere and
should also be taken into account (\citealp{LUNetal2010}).

\section{Discussion}

Observational evidence has been provided for examples of sausage
oscillations in magnetic pores with a range of periods from $\sim50$
to $\sim600$~s. To our knowledge, this is only the second reported
observation of oscillatory behavior in the area of a magnetic
waveguide after \cite{DORetal2008}. However, here we have a higher
cadence allowing wave phenomena occurring on much shorter
time-scales to be resolved. Our observations also provide intensity
information about the interior plasma of the pore, where oscillatory
behavior is also seen.

The oscillatory phenomena were identified using a relatively new
technique (at least to solar applications) known as Empirical Mode
Decomposition. Direct comparison between intensity IMFs and pore
size IMFs allowed the identification of out of phase oscillatory
signatures, indicating the presence of sausage oscillations. This
was supported by calculation of the relative phase between the two
time series. This method highlights the difficulties in obtaining a
clear signal of sausage oscillations by measuring the
cross-sectional area of the waveguide. Periodic changes in pore size
can be related to periodic behavior in intensity (i.e. in phase
periodic behavior with similar power). Using a method like wavelet
analysis alone would not allow the comparison of the phases of
intensity and pore size without ambiguity. Hence, wavelet analysis
alone could lead to false detections of sausage modes.

 Only a few sections in the various IMFs provided a clear
signal of sausage oscillations, were the oscillation in pore size
and intensity in the pore are out of phase by $180$ degrees. The
other signals in Table $1$ show oscillations in pore size that are
not identified with a change in intensity, hence, are a somewhat
weaker signal of sausage oscillation. The periodic phenomena in
intensity can most likely be attributed to global acoustic modes
which are omnipresent in the photosphere and are found both inside
the pore and in the surrounding granules and have a wide range of
periods from tens of second to tens of minutes (see, e.g.
\citealp{JESetal2007}).

Due to the omnipresent nature of the global acoustic oscillation in
the surrounding granules, it is proposed the main driver for these
oscillations is the global acoustic mode. The periods found mainly
correspond to the $3$ and $5$ minute global modes that have been
numerously reported, adding further strength to the argument. The
question of whether the detected modes are standing or propagating
remains unanswered due to the lack of velocity data, which could
have answered this. For example, the signals at $t=1700$~s in $c_3$
and $c_4$ could correspond to different harmonics of the standing
sausage oscillation. A plausible suggestion that the oscillations
were standing modes led to the estimation of waveguide lengths that
are comparable to the distance from the photosphere to the
transition region, i.e. the two boundaries required to set up
standing modes in the lower solar atmosphere.

\begin{acknowledgements}
The authors thank J. Terradas for providing the EMD routines used in
the data analysis. RE acknowledges M. K\'eray for patient
encouragement. The authors are also grateful to NSF, Hungary (OTKA,
Ref. No. K67746) and the Science and Technology Facilities Council
(STFC), UK for the financial support they received. Observations
were obtained at the National Solar Observatory, operated by the
Association of Universities for Research in Astronomy, Inc. (AURA),
under agreement with the National Science Foundation. We would like
to thank the technical staff at DST for their help and support
during the observations.
\end{acknowledgements}

\end{document}